\documentclass[report]{amsart}
\usepackage{amsmath, amsthm, amsfonts, amssymb, epic, ecltree}
\usepackage{graphicx}
\usepackage{ifpdf}
\usepackage{color}
\usepackage{setspace}

\usepackage{epic,bez123}
\usepackage{wrapfig}

\DeclareGraphicsExtensions{.eps,.bmp}

\newtheorem*{rf}{Relative Frequency}
\newtheorem*{rand}{Randomness}

\title{Typical Worlds}

\author{Jeffrey A. Barrett}

\date{}

\begin{document}

\maketitle

\begin{abstract}
Hugh Everett III presented pure wave mechanics, sometimes referred to as the many-worlds interpretation, as a solution to the quantum measurement problem. While pure wave mechanics is an objectively deterministic physical theory with no probabilities, Everett sought to show how the theory might be understood as making the standard quantum statistical predictions as appearances to observers who were themselves described by the theory. We will consider his argument and how it depends on a particular notion of branch typicality. We will also consider responses to Everett and the relationship between typicality and probability. The suggestion will be that pure wave mechanics requires a number of significant auxiliary assumptions in order to make anything like the standard quantum predictions.
\end{abstract}

\section{introduction}

It is tempting to claim that one can derive strong conclusions from weak assumptions, but there is a no-magic constraint. Since the conclusion of a valid deductive argument cannot be stronger than the premises required to get it, if one gets a strong conclusion, one must have made similarly strong assumptions somewhere along the way. In practice, however, it is often difficult to say where.


Hugh Everett III (1955a, 1955b, 1956, 1957) presented pure wave mechanics, also known as the many-worlds interpretation, as a solution to the quantum measurement problem.\footnote{Everett himself did not refer to branches as worlds in his written work. See Barrett (2011b, 2016a) for discussions of his metaphysical commitments and the role of metaphysics in interpreting pure wave mechanics more generally.} While he believed that there was a sense in which one could derive the standard quantum statistics from pure wave mechanics alone, his approach was significantly less ambitious than deducing probabilistic predictions. That said, even Everett's relative weak account of quantum statistics requires significant additions to pure wave mechanics.

Here we are concerned with Everett's presentation of pure wave mechanics, his derivation of the quantum statistics, and DeWitt's and Graham's response. We will pay particular attention to the explanatory role played by alternative notions of branch typicality and the relationship between typicality and probability. The aim is to understand Everett's approach better and to get a sense of the strength of the auxiliary assumptions one would need to derive anything like the standard quantum probabilities from pure wave mechanics.\footnote{See Barrett (2016b) for a preliminary discussion of typicality in pure wave mechanics.}

The present paper seeks to get at what Everett thought about typicality and probability, including his reaction to DeWitt and Graham's criticisms. While there are methodological morals to this story that clearly apply more generally, other proposals for how to understand quantum typicality and probability require careful analysis on their own terms.\footnote{Questions concerning the relationship between typicality and probability arise in other formulations of quantum mechanics as well. For a recent discussion of typicality and probability in Bohmian mechanics see Goldstein (2012).} The present argument involves how one might best identify a physical theory for the purpose of probabilistic explanation. Since it is implausible that there are canonical criteria for how to individuate theories or for what constitutes a good explanation, the approach will be thoroughly pragmatic. How the methodological morals might be best applied beyond the story told here is largely left to the reader.

\section{pure wave mechanics}

While pure wave mechanics is an objectively deterministic theory that says nothing whatsoever about probability, Everett sought to describe a sense in which it might be understood as making the same statistical predictions as the standard von Neumann-Dirac collapse theory.\footnote{See Dirac (1930) and von Neumann (1932, 1955) for early descriptions of the standard collapse theory.} The thought was to get the standard quantum predictions as subjective appearances to observers who were themselves described quantum-mechanically. What this meant for Everett was that the relative measurement records of a \emph{typical} relative observer will exhibit the standard quantum statistics. In order to track the various auxiliary assumptions required, we will start with a specification of pure wave mechanics.

Everett presented pure wave mechanics as a modification of the standard collapse formulation of quantum mechanics. And, following von Neumann (1955), Everett took the standard collapse formulation of quantum mechanics to be characterized by the following principles:

\begin{itemize}

\item[1.] {Representation of States}: The state of a physical system $S$ is represented by a vector $|\psi\rangle_S$ of unit length in a Hilbert space $\mathcal{H}$.

\item[2.] {Representation of Observables}: A physical observable $O$ is represented by a set of orthogonal vectors $\mathcal{O}$. These vectors represent the eigenstates of the observable, each corresponding to a different value.

\item[3.] {Dynamical Laws}:

\item[] I. {Linear dynamics}: If \emph{no measurement} is made, the system $S$ evolves in a deterministic linear way: $|\psi(t_1)\rangle_S=\hat{U}(t_0,t_1)|\psi(t_0)\rangle_S$.

\item[] II. {Nonlinear collapse dynamics}: If a \emph{measurement} is made, the system $S$ randomly, instantaneously, and nonlinearly jumps to an eigenstate of the observable being measured: the probability of jumping to $|\phi\rangle_S$ when $O$ is measured is~$|\langle\psi|\phi\rangle|^2$.

\end{itemize}

In addition to these principles, Everett appealed to the standard eigenvalue-eigenstate link to attribute \emph{absolute} properties to a system. In particular, a system $S$ has an \emph{absolute} value for observable $O$ if and only if $|\psi\rangle_S \in \mathcal{O}$, and the value is given by the eigenvalue corresponding to~$|\psi\rangle_S$.

Everett believed, however, that its incompatible dynamical laws rendered the von Neumann-Dirac collapse theory logically inconsistent and hence untenable. He presented what he called the \emph{question of the consistency} of the standard theory in the context of an ``amusing, but \emph{extremely hypothetical}~drama'' (1956, 74).\footnote{Everett was arguably too critical here. The theory as it stands is just ambiguous inasmuch as one does not know precisely when to apply rules~3I and~3II. That is, one could remove the threat of inconsistency by specifying strictly disjoint conditions for when each rule obtains. This is what Wigner later sought to do. For his part, Everett assumed that a measuring device should evolve linearly like every other physical system. It is this auxiliary assumption that yields the inconsistency.} This would later become known as the Wigner's Friend story after Eugene Wigner (1961) retold it without attribution to Everett.

A version of the story goes as follows. Suppose that a spin-$1/2$ system $S$ begins in the state
\begin{equation}
\label{eq1}\alpha |\!\uparrow_x\rangle_S + \beta |\!\downarrow_x\rangle_S
\end{equation}
and a friend $F$ and his measuring device $M$ begin ready to make a measurement of the $x$-spin of $S$. Assuming perfect correlating interactions, the linear dynamics (von Neumann's Process~2, rule~3II above) predicts that the resultant state will be:

\begin{equation}
\label{eq2}\alpha |\mbox{``$\uparrow_x$''}\rangle_F |\mbox{``$\uparrow_x$''}\rangle_M |\!\uparrow_x\rangle_S + \beta |\mbox{``$\downarrow_x$''}\rangle_F |\mbox{``$\downarrow_x$''}\rangle_M |\!\downarrow_x\rangle_S.
\end{equation}

In contrast, if one were to suppose, as suggested by the standard collapse theory, that there is something special about the friend or his measuring device that causes a collapse of the state of his object system on measurement, one would end up with one of the states predicted by the collapse dynamics (von Neumann's Process~1, rule~3I above):

\begin{equation}
\label{eq3}|\mbox{``$\uparrow_x$''}\rangle_F |\mbox{``$\uparrow_x$''}\rangle_M |\!\uparrow_x\rangle_S
\end{equation}
or

\begin{equation}
\label{eq4}|\mbox{``$\downarrow_x$''}\rangle_F |\mbox{``$\downarrow_x$''}\rangle_M |\!\downarrow_x\rangle_S
\end{equation}
with probabilities $|\alpha|^2$ and $|\beta|^2$~respectively.

The moral of the story is that the two dynamical laws of the standard von-Neumann-Dirac formulation of quantum mechanics predict incompatible states when applied to the same physical interaction. Further, as Everett explicitly recognized, state~(\ref{eq2}) is in principle empirically distinguishable from states~(\ref{eq3}) and~(\ref{eq4}) by an inference measurement on the composite system consisting of $F$, $M$, and $S$.\footnote{This might be accomplished by measuring an observable that has state~(\ref{eq2}) and the orthogonal state one gets by subtracting the second term rather than adding it to the first as eigenstates corresponding to different eigenvalues.}

Everett held that one only has a satisfactory formulation of quantum mechanics if one can provide a satisfactory account of such nested measurements. That is, if one cannot tell the Wigner's Friend story consistently, then one does not have a satisfactory formulation of quantum mechanics.

Everett believed that the Wigner's Friend story could be told simply and consistently in the context of pure wave mechanics, the theory one gets by starting with the standard collapse theory and simply deleting the collapse dynamics~(rule~3II). In particular, he believed that the final state after~$F$'s measurement interaction is simply given by state~(\ref{eq2}), thus removing the possibility of a contradiction. And, again, he believed that an external observer would in principle be able to show this empirically by an appropriate interference measurement on the composite system $F$, $M$, and~$S$.\footnote{In other words, the linear dynamics entails that Everett worlds cannot be causally closed. This point is also discussed in Albert (1986) and Albert and Barrett (1995), and a point that we will return to later.} But he also believed that it will appear to~$F$ that she has a perfectly determinate measurement outcome, and, more generally, that one can recover the standard quantum statistics for such appearances from pure wave mechanics alone. He took pure wave mechanics thereby to provide a satisfactory resolution to the quantum measurement problem.

As Everett described his project, ``we shall deduce the probabilistic assertions of [the collapse dynamics (3II)] as \emph{subjective} appearances'' to observers who are themselves treated as perfectly ordinary physical systems always subject to the linear dynamics (3I) ``thus placing the theory in correspondence with experience.'' The upshot is that ``We are then led to the novel situation in which the formal theory is objectively continuous and causal, while subjectively discontinuous and probabilistic.'' Everett took this to solve the nested measurement problem because ``while this point of view thus shall ultimately justify our use of the statistical assertions of the orthodox view, it enables us to do so in a logically consistent manner, allowing for the existence of other observers'' (1956, 77--8). Specifically, this amounted to describing a sense in which pure wave mechanics predicted the standard quantum statistics for the relative measurement records of a typical relative observer.\footnote{There is a long tradition of physicists and philosophers who have sought as Everett did to deduce the standard quantum probabilities from pure wave mechanics. The list includes, among others, Hartle (1968), DeWitt (1971), Graham (1973), Farhi, Goldstone and Gutmann (1989), Deutsch (1999), Zurek (2005), Saunders (2010b), Wallace (2010b, 2012), and Sebens and Carroll (2016).  While the details of the arguments and precisely what is meant by quantum probability varies significantly, each of these deductions relies on auxiliary assumptions that go beyond pure wave mechanics, at least as Everett understood the theory.}

Everett's first step in deriving the standard quantum statistics as typical was to introduce a distinction between relative and absolute states. This distinction plays an essential explanatory role in his relative-state formulation of pure wave mechanics. While pure wave mechanics does not describe an observer as having any particular \emph{absolute record} after a typical measurement interaction, it does describe the observer as having a number of different \emph{relative records}. In the case of state~(\ref{eq2}), there exists, as Everett put it, a ``cross section'' of the total state where each term describes a branch where both the friend and the measuring device have determinate relative measurement records (1955a, 66--8). Namely, in the determinate record basis, one term describes $F$ as getting the relative result ``$\uparrow_x$'' (relative to $S$ being $x$-spin up) and the other describes $F$ as getting the relative result ``$\downarrow_x$'' (relative to $S$ being $x$-spin down). Everett took the existence of branches where an observer has determinate relative measurement records to explain the observer's determinate~experience. As he put it to Abner Shimony some years later, ``Each individual branch looks like a perfectly respectable world where definite things have happened'' (Barrett and Byrne 2012, 275--6).\footnote{How things will \emph{look} to a relative observer was for Everett a matter of the sequence of relative records that the observer would have under the linear dynamics. Everett's aim was to argue that a typical relative observer will have records that are distributed the same way as the measurement records we take ourselves to have. It is in this sense that he believed he could show that a typical relative observer's world would look perfectly ordinary.}

The sense in which Everett thought that each branch would look like a perfectly respectable world requires care to sort out.\footnote{See Barrett (1999, 2011a, 2011b, 2015) for discussions.}  For present purposes, we will simply suppose that one can understand an observer's experience as supervening on the relative records characterized by a single branch in a determinate record basis for that observer and turn to consider his deduction of the standard quantum statistics.

\section{Everett's deduction}

In the Wigner's friend story, the standard collapse theory predicts states~(\ref{eq3}) and~(\ref{eq4}) after the measurement with probabilities $|\alpha|^2$ and $|\beta|^2$ respectively. The distribution of results one gets from a random process with these probabilities represents the statistics Everett wanted to recapture in pure wave mechanics. The problem is that pure wave mechanics simply predicts that the post-measurement state is~(\ref{eq2}). Further, it predicts no stochastic events and there is no epistemic uncertainty regarding what the absolute and relative states are after the interaction. Hence pure wave mechanics alone makes no probabilistic predictions whatsoever. It just tells what the final state will be, and the final state is not one where either possible outcome has been realized at the exclusion of the other. Hence there is not even a particular outcome for which $|\alpha|^2$ or $|\beta|^2$ might have been the probabilities before the measurement.

For his part, Everett clearly and repeatedly insisted that there were no probabilities in pure wave mechanics, a view that is reflected in the original title of his PhD thesis ``Wave Mechanics without Probability'' (Barrett and Byrne 2012, 72). Rather than claim that his theory predicted the same chance events as predicted by the collapse dynamics (von Neumann's Process~1, rule~3II above), Everett argued that it would appear to a typical relative observer that there had been such events. More specifically, he argued that an observer's relative measurement records in a \emph{typical branch} would be randomly distributed with the standard quantum probabilities, in the measure of typicality provided by the norm-squared of the coefficient associated with each branch in a determinate-record expansion of the total state. 

The notion of a typical branch played an essential role in Everett's account of the standard quantum statistics. He took finding an appropriate typicality measure for branches in pure wave mechanics to be analogous to the problem of finding an appropriate typicality measure for states in statistical mechanics.
\begin{quote}
In order to establish quantitative results, we must put some sort of measure (weighting) on the elements of a final superposition. This is necessary to be able to make assertions which hold for almost all of the observer states described by elements of a superposition. We wish to make quantitative statements about the relative frequencies of the different possible results of observation---which are recorded in the memory---for a typical observer state; but to accomplish this we must have a method for selecting a typical element from a superposition of orthogonal states'' (1956, 123--4; 1957, 190).
\end{quote}
We will return to the analogy with statistical mechanics later.

Everett's strategy was to impose a sequence of desirable constraints until a particular typicality measure over branches was uniquely determined. Importantly, he did not argue that almost all branches \emph{by count} will exhibit the standard quantum statistics. Rather, he argued that almost all branches \emph{in a natural measure~$m$ given the mathematical structure of pure wave mechanics} will exhibit the standard quantum statistics.

While he clearly and repeatedly reminded the reader that his typicality measure~$m$ did not represent probabilities, he also explicitly took advantage of the fact that it satisfied the formal conditions for being a probability measure over the set of branches~$B$ determined by a specified orthogonal decomposition of the state (1956, 79--80, 127). In particular, the set function $m$ assigns a number between zero and one to each subset of $B$ such that (1)~$m(B) = 1$, (2)~$m(Q) = 1- m(\bar{Q})$ for subset~$Q$, (3)~if~$R \cap Q = \emptyset$, then~$m(R \cup Q) = m(R) + m(Q)$ for subsets~$Q$ and~$R$, and (4)~$m$ is countably additive for countable unions of disjoint subsets of~$B$.

Beyond this, Everett's first constraint was that~$m$ be a positive function of the complex-valued coefficients~$a_i$ associated with the branches of the superposition~$\sum a_i |\chi_i\rangle$. This may seem natural enough inasmuch as, short of just counting the branches and considering relative proportions, one might argue that pure wave mechanics itself does not provide much else that one might use as a typicality measure over branches.\footnote{But even this first move of requiring that one's typicality measure be a function of branch amplitudes is in no way forced on us by the theory. As David Albert (2010, 360) argues, one could take one's typicality measure to be a function of the \emph{basis elements} describing the branches. Albert's specific suggestion is that one might take the most natural measure over branches to be a function of how fat one is in each branch if that is what one cares about most.}

That said, there are many functions of the coefficients~$a_i$ one might consider. His second constraint addressed this in part by requiring that~$m$ be a function of the amplitudes of the coefficients alone. The reason he gave was that the coefficients on branches $a_i$ can only be empirically determined up to an arbitrary phase factor.

Finally, he stipulated a sub-branch additivity condition. Since any collection of branches $b_i$ of the total state can be considered to be a single branch $b^*$ on a different orthogonal decomposition of the state, he required that the measure assigned to $b^*$ be equal to the sum of the measures assigned to the branches $b_i$. Synchronically, this provides a natural nesting relation between the typicality measures assigned to branches on different cross~sections. Diachronically, it ensures a conservation of typicality of branches under the linear dynamics that Everett found attractive. Regarding this diachronic consideration, he explained,
\begin{quote}
[W]e wish to make statements about `trajectories' of observers. However, for us a trajectory is constantly branching (transforming from state to superposition) with each successive measurement. To have a requirement analogous to the `conservation of probability' in the classical case, we demand that the measure assigned to a trajectory at one time shall equal the sum of the measures of its separate branches at a later time. This is precisely the additivity requirement which we imposed and which leads uniquely to the choice of square-amplitude measure. Our procedure is therefore quite as justified as that of classical statistical~mechanics (1956, 126; 1957, 192).
\end{quote}
Insofar as Everett had in mind an analogy between the conservation of probability over non-splitting trajectories in classical statistical mechanics as described by Liouville's theorem (which he explicitly mentions here) and the conservation of typicality over splitting branches in pure wave mechanics, it is unclear precisely how he intended for it to go. What he said is that such a conservation condition is ``the only choice which makes possible any reasonable statistical deductions at all'' (1956, 125; 1957, 192). Among the diachronic deductions one can get given the sub-branch additivity condition is that, given a specified initial branch where a measurement interaction occurs~$b$, if one assumes no interference between descendent branches, then the typicalities associated with the descendent branches behave like a probability measure conditional on the state of~$b$. One need not require that a typicality measure have this property, but it is unsurprising that sub-branch additivity seemed natural to Everett.

Together, he took the three stipulated constraints together with the properties of a probability measure to determine uniquely the typicality measure~$m$ as the norm-squared-amplitude measure on branches.\footnote{Everett originally just wanted to argue that his choice of typicality measure was ``not as arbitrary as it appears.'' Some years later, when editing his thesis for inclusion in the DeWitt and Graham (1973) anthology, he changed the text to read instead ``there is a unique measure which will satisfy our requirements'' Barrett and Byrne (2012, 359).}

It was important to Everett that~$m$ did not represent branch \emph{probabilities}. Indeed, inasmuch as all branches are equally actual, he took the probability of each branch to be one. And since they were each in principle observable, they were actual.
\begin{quote}
Take this opportunity to caution against a certain viewpoint which can lead to difficulties. This is the idea that, after an apparatus has contracted with the system, in `actualityÕ one or another of the elements of the resultant superposition described by the composite state-function has been realized to the exclusion of the rest, the existing one simply being unknown to the external observer \ldots. This position must be erroneous since there is always the possibility for the external observer to make use of interference properties between elements of the superposition. (1956, 149)
\end{quote}
There is, then, no actualization of one measurement outcome at the exclusion of the rest. They are all equally actual.

This point is closely tied to Everett's view that in the Wigner's Friend story the external observer may always in principle perform a measurement that would show that the system containing the friend~$F$ is in the superposition~(\ref{eq2}) predicted by the linear dynamics. All branches are equally actual precisely because Everett believed that it was always in principle possible to observe interference effects between them. As he put the point, ÒIt is \ldots improper to attribute any less validity or ÔrealityÕ to any element
of a superposition than any other element, due to [the] ever present possibility of obtaining interference effects between the elements. All elements of a superposition
must be regarded as simultaneously existingÓ (1956, 150). And it is a direct consequence of this that, contrary to the understanding of many Everettians, the reality of branches for Everett did not in any way depend on the selection of a preferred basis or on special decoherence conditions obtaining.\footnote{See Wallace (2010a, 69--71) for a recent reading of the metaphysical role of decoherence in the Everett interpretation.}  Indeed, decoherence only serves to make it more difficult to show empirically that all the branches predicted by the linear dynamics are in fact equally actual.

Returning to his choice of typicality measure, while~$m$ may seem a natural measure given the structure of pure wave mechanics, it is clearly not the only measure Everett might have chosen. Indeed, as we will see in the next section, some of Everett's strongest supporters believed that he had chosen the wrong measure. While there are more and less natural candidates, there is no canonical way to assign a measure over branches. Hence, Everett's choice of typicality measure is an addition to the theory.

For his part, Everett took measure~$m$ to be necessary for talking sensibly about the statistical properties of \emph{typical} branches. The main result of his thesis was that almost all branches in measure~$m$ will exhibit the standard quantum statistics in the limit as the number of measurements one has performed gets large. This was the fact that he took to be most significant in understanding pure wave mechanics.\footnote{See his discussion of this at the Xavier conference. Here Everett again describes the analogy with classical statistical mechanics (Barrett and Byrne 2012, 274--5).}

Consider a measuring device $M$ that is ready to make and record the results of an infinite series of measurements on each of an infinite series of systems $S_k$ in initial~state
\begin{equation} 
\alpha |\!\uparrow_x\rangle_{S_k} + \beta |\!\downarrow_x\rangle_{S_k}.
\end{equation}
Suppose that $M$ interacts with each system in turn and perfectly correlates its $k$th memory register with the $x$-spin of system $S_k$ by the linear~dynamics. This will produce an increasingly complicated entangled superposition of different sequences of measurement outcomes. After one measurement, the state of $M$ and $S_1$ in the determinate record basis will be
\begin{equation}
\alpha |\mbox{``$\uparrow$''}\rangle_M |\!\uparrow\rangle_{S_1} + \beta |\mbox{``$\downarrow$''}\rangle_M |\!\downarrow\rangle_{S_1}.
\end{equation}
After two measurements, the state of $M$, $S_1$, and $S_2$ will be
\begin{equation}
\begin{aligned}
\alpha^2 |\mbox{``$\uparrow\uparrow$''}\rangle_M |\!\uparrow\rangle_{S_1} |\!\uparrow\rangle_{S_2}
&+ \alpha\beta |\mbox{``$\uparrow\downarrow$''}\rangle_M |\!\uparrow\rangle_{S_1} |\!\downarrow\rangle_{S_2} \\
&+ \beta\alpha |\mbox{``$\downarrow\uparrow$''}\rangle_M |\!\downarrow\rangle_{S_1} |\!\uparrow\rangle_{S_2}
+ \beta^2 |\mbox{``$\downarrow\downarrow$''}\rangle_M |\!\downarrow\rangle_{S_1} |\!\downarrow_x\rangle_{S_2}.
\end{aligned}
\end{equation}
Figure~1 illustrates the first three steps of the branching process and the amplitudes associated with each branch.
\begin{figure}[htb]    \centering
\includegraphics[width=4.0in]{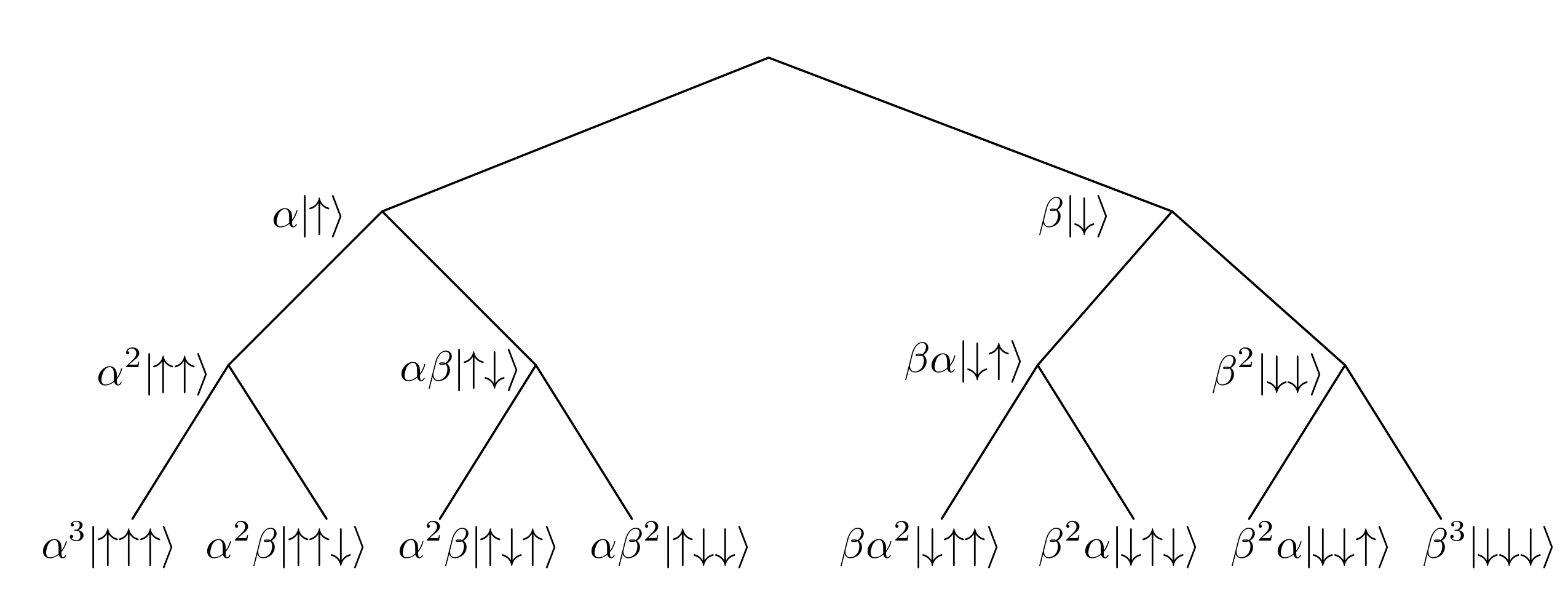}
\end{figure}

Using this setup, Everett (1956, 127) provided a rough sketch of the proof of the following theorem :
\begin{rf}
For any $\delta>0$ and $\epsilon>0$, there exists a $k$ such that after $k$ measurements the sum of the norm-squared of the amplitude associated with each branch where the distribution of spin-up results is within $\epsilon$ of $|\alpha|^2$ and the distribution of spin-down results is within $\epsilon$ of $|\beta|^2$ is within $\delta$ of~one.
\end{rf}
And he hinted at something like the following (1956, 127--8):
\begin{rand}
The sum of the norm-squared of the amplitude associated with each branch where the sequence of relative records is algorithmically incompressible, or satisfies any other of the standard criteria for being random, goes to one as the number of measurements gets~large.
\end{rand}

Both of these results can be proven for pure wave mechanics with modest auxiliary assumptions.\footnote{See Barrett (1999) for a discussion of the sort of assumptions one needs regarding the properties of limiting states to get such results and Albert (2010, 357) for a recent discussion of the theorem.} In this sense, it is fair to conclude with Everett that in the limit as the number of measurements gets large, almost all branches in measure~$m$ will describe sequences of measurement records that are randomly distributed with the standard quantum statistics.

He then proceeded to argue that one should expect a similar result to hold for any situation where one is performing a series of repeated quantum measurements regardless of whether they produced perfect correlations or concerned repeated measurements of the same observable (1956, 129--30). He thus concluded that the relative measurement records of a typical relative observer will exhibit the standard quantum statistics in the limit. Hence it would appear to the typical relative observer that the standard quantum probabilities obtain for the results of her measurements. This was Everett's deduction of the standard quantum statistics from pure wave mechanics.

\section{Indifference and a new partition}

Everett held that our quantum experience is fully explained by the fact that the relative records of a typical relative observer, in the sense of typical specified, will exhibit the standard quantum statistics. But even many his supporters found this explanation unsatisfactory.

Bryce DeWitt (1971) and Neill Graham (1973) argued that Everett's choice of typicality measure was unmotivated. Their worry was that it is typically not the case that the records of most relative observers \emph{by count} will exhibit the standard quantum statistics. As Graham put it, ``it is extremely difficult to see what significance such a measure can have when its implications are completely contradicted by a simple count of the worlds involved'' (1973, 236).

What DeWitt and Graham wanted was a theory where most worlds by count exhibit the standard quantum statistics. Only then did they believe that one would have a satisfactory explanation for why one should expect one's experiments to exhibit the standard quantum statistics. To this end, Graham stipulated a rule for how worlds split such that the number of worlds that exhibit a particular outcome after a measurement interaction was proportional to the square of the coefficient on the term describing that outcome. This ad hoc splitting procedure ensured that almost all worlds by simple count will exhibit the standard quantum statistics.\footnote{Graham presented his ``two step'' splitting procedure as an extension of what he called Everett's ``one step'' account of measurement. On Graham's account, one introduces a third macroscopic apparatus that mediates between a microscopic system and a macroscopic observer, then comes to thermodynamically equilibrium with its environment. The partition of branches that he stipulated for the resulting state differs from the determinate-record partition that Everett derived by assuming that the linear dynamics fully described the measurement interaction.}

We know what Everett thought of DeWitt's criticism and Graham's proposal from his marginal notes on his personal copy of the DeWitt-Graham (1973) anthology. Where DeWitt (1971, 185) claimed that Everett's typicality argument was unsatisfying, Everett wrote in the margin ``only to \underline{you}!'' And where Graham (1973, 236) claimed that Everett's typicality measure was unmotivated, Everett wrote in the margin ``bullshit."\footnote{See Barrett and Byrne (2012, 365--6) for photocopies of these pages.}

Everett was certainly free to use any measure he wanted---it just needed to be a measure that allowed him to argue that the relative records in a typical branch will exhibit the standard quantum statistics. Further, as we have seen, the measure he used, while in no way canonical, has a natural feel to it given the formal structure of pure wave mechanics.

That DeWitt and Graham believed that making something true of most branches by count was the only way to explain why an observer should expect his results to exhibit the standard quantum statistics tells us something about their background commitments. To begin, they took the talk of worlds somehow inhabited by copies of the observer seriously, arguably much more seriously than Everett himself did.\footnote{See Barrett (2011b) for a discussion of Everett's metaphysical views.} Second, they seem to have wanted standard probabilistic predictions from the theory, something that Everett did not seek to provide. And finally, their intuitions were apparently guided by a \emph{principle of indifference}, the sense that if there are $n$~possibilities, then, in the absence of other information, one should assign probability~$1/n$ to each possibility.\footnote{There are a number of discussions of difficulties one encounters in applying the principle of indifference to branches in the literature. See, for examples, Barrett (1999, 168--73) and, more recently, Greaves (2007).}

An account of quantum probabilities along these lines might go something like this. The overwhelming majority of quantum worlds by simple count exhibit the standard quantum statistics on the new stipulated partition. Since each world is equally likely, one should thus with high probability expect to find that one's actual measurement results exhibit the standard quantum statistics.

There are, however, several problems with this line of argument. First, it is unclear precisely what such quantum probabilities are probabilities of. Since there are no chance events nor uncertainties stipulated by pure wave mechanics, the theory itself does not provide an candidate for what they might refer to. Even if one adds a preferred partition of worlds, inasmuch as there is a copy of the observer in each of the post-measurement worlds, one would need to explain why the probability of finding the observer in any particular world is something other than one. One option is to try to make sense of self-location probabilities. The idea here would be that since most quantum worlds by simple count exhibit the standard quantum statistics on Graham's partition, one should with high probability expect \emph{to find one's self} in a world where one's measurement results exhibit the standard quantum statistics. This strategy would require one to say both how one assigns probabilities and to what one assigns them, which means making substantive additions to the theory. If this is what they had in mind, neither DeWitt nor Graham explain how to do it.\footnote{Self-location probabilities are a recurring topic in Saunders, Barrett, Kent, and Wallace (eds) (2010). See also Vaidman (2012). Albert and Loewer's (1988) single- and many-minds theories illustrate the sort of assumptions regarding probabilities and what they concern that one would need to add to pure wave mechanics to make sense of self-location probabilities. We will briefly discuss these theories later.}

A second problem is that the content of a principle of indifference always depends on one's choice of a partition. Bas van Fraassen's (1989) cube factory story illustrates the point. Suppose one only knows that a factory produces cubes with a side between~0 and~2 meters. If one considers side-length, one might imagine that a principle of indifference requires that one take the probability of a randomly selected cube having a side between~0 and~1 to be~$1/2$ since side-lengths range from~0 to~2. If one considers volume, one might imagine that a principle of indifference requires that one take the probability of a randomly selected cube having a volume between~0 and~1 to be~$1/4$ since volumes range from~0 to~4. But since having a side length between~0 and~1 is the same thing as having a volume between~0 and~1, the different partitions, each perfectly natural given different interests, yield inconsistent probabilities. The upshot is that the content of a principle of indifference depends on one's partition of a space of possibilities, different partitions yield inconsistent probability assignments, and what partitions one finds natural often just depend on one's interests. For this reason, applications of the principle of indifference are often clearly ad hoc given the proposed partition.

Since pure wave mechanics specifies no canonical decomposition of the full state, it does not tell one what set of branches one should use in assigning each branch probability~$1/n$.\footnote{Indifferent priors are sometimes suggested as a way of representing a complete lack of initial information. But judgments of lack of information are also relative to one's interests. If one is interested in information regarding side length, one partition may seem more natural; if one is interested in information regarding volume, another partition may. Similarly, in the context of pure wave mechanics, if one is interested in information regarding determinate measurement records, one decomposition of the absolute state may seem more natural; if one is interested in information regarding energy another decomposition may. And, as we just saw, the application of a principle of indifference over different partitions typically yields incoherent probability assignments.} This problem is made particularly salient by the fact that choosing a determinate-record basis in order to guarantee that each relative observer has determinate measurement records then assigning each branch probability~$1/n$ typically yields the wrong quantum probabilities. DeWitt and Graham must assume both a special, preferred way to individuate branches and a principle of indifference even to get started in deriving the standard quantum probabilities. Further, stipulating precisely the right number of duplicate worlds to get the right quantum probabilities on a principle of indifference is manifestly ad hoc.

It is important to be clear on this point. If one individuates branches in the determinate-record basic as Everett did, supposing that there is one branch for each different relative sequence of measurement results then assigns unbiased priors by appeal to a principle of indifference, one gets the wrong quantum probabilities. Hence even an appeal to a standard sort of unbiased principle of indifference is not enough to get the standard quantum probabilities. If one wants to ground one's probability assignment in a principle of indifference one must either adjust how one individuates branches or adopt an exotic principle of indifference that provides the standard probabilities to each branch as Everett individuated them.\footnote{Sebens and Carroll (2016) pursue this second strategy. They argue that while ``it is tempting to regard each branch as equiprobable \ldots the temptation should be resisted.'' Rather, they believe that assigning the standard quantum probabilities to each branch constitutes ``the uniquely rational way of apportioning credence in Everettian quantum mechanics.'' To get this, they appeal to something they call the \emph{Strong Epistemic Separability Principle}, a new basic principle of reason that they take to be a generalized principle of indifference.}

If one stipulates that the simple proportion of branches with a particular property is equal to the quantum probabilities for the property, then applies a principle of indifference, one will clearly get the standard quantum predictions. But there is no mystery about what does the work here. One gets the right predictions out because one put them in by stipulating how to count worlds. Similarly, if one individuates branches like Everett did but stipulates that the probability of each branch is equal to the norm-squared coefficient associated with the branch, one will also get the standard quantum predictions. And here there is even less of a mystery. One gets the right predictions out because one is simply assigning the standard quantum probabilities to each branch. That DeWitt and Graham disagreed with Everett concerning both the appropriate partition and the appropriate measure is evidence that neither is canonical, and that both stipulations nevertheless yield the standard quantum measure over branches is evidence of each being ad hoc. The upshot is that one can get the standard quantum probabilities by making an ad hoc choice for how branches are individuated and/or an ad hoc choice for one's probability measure over branches.

That one's partition and probability measure necessarily work together has implications for the status of a principle of indifference. Inasmuch as the directive to assign probabilities of~$1/n$ to each of~$n$ possibilities does not mean anything apart from a choice of partition, it cannot be a basic principle of reason. Further, insofar as one believes that there is no canonical way of partitioning possibilities that applies over all domains of application, no principle-of-indifference/partition pair can be adopted as a basic principle of reason either. The next step might be to imagine that the basic principle of reason consists in a principle-of-indifference/partition/domain-of-application triple, but it is difficult to imagine how a three-parameter principle of indifference could fail to look ad hoc.

To be sure, people often claim that there are canonical partitions and canonical probability measures over those partitions even as they disagree with each other regarding what the canonical partitions and measures are. One need not look far for a plausible diagnosis. An inquirer might assign any set of coherent, non-dogmatic priors to the~$n$ hypotheses she is considering without fear of finding herself committed to a Dutch book or failing to respond appropriately to relevant evidence. That a rational inquirer enjoys this freedom illustrates why there can be no basic principle of reason requiring any particular assignment of priors. She might assign a probability of~$1/n$ to each hypothesis, but only if the partition represented by her hypotheses in fact yields unbiased degrees of belief given her actual commitments. Otherwise, she \emph{would} be open to a Dutch book. In short, a rational inquirer must assign those credences to which she finds herself currently committed to each of her~$n$ hypotheses, whatever they may be. Of course, she may take a partition where each hypothesis gets the same probability given her credences to be particularly natural. If so, what may look like an application of a principle of indifference is just a result of the choice of a symmetric partition given the agent's prior epistemic commitments.\footnote{See Barrett (2014) for a signaling-game account of how a descriptive partition and an assignment of effective priors might coevolve. The model illustrates how a rich set of effective priors might evolve by means of precisely the same process by which one comes to have a reliable descriptive language. Such effective priors may evolve to be unbiased over the evolved descriptions on this sort of model if agents are, for some reason, rewarded for using maximally informative signals.}

It is unsurprising that we find those partitions and measures that yield the empirically warranted probabilities natural. But if one wants the standard quantum probabilities from pure wave mechanics, one needs significant auxiliary assumptions. And, given that pure wave mechanics does nothing to constrain even a measure of typicality over branches, such assumptions are not well-characterized as necessary constraints on reason.

While this is a different sort of problem, it is also worth noting a technical issue that arises when one tries to apply a principle of indifference in the case of pure wave mechanics. If one insists that there are only a finite number of worlds, then a principle of indifference cannot provide precisely the same fine-grained quantum statistics as predicted by the standard theory. But if one allows a countably infinite number of worlds and assumes sigma-additivity for the measure over those worlds (assumption~4 in the list of conditions above), then there can be no unbiased priors at all. In this case, a principle of indifference does not even make sense.\footnote{It is presumably for this reason that both Graham (1973, 251--2) and Sebens and Carroll (2016) require the number of Everett worlds be finite.}

For his part, Everett had no use for a principle of indifference inasmuch as he did not want or need probabilities over branches. Rather, his account of quantum probabilities concerned \emph{what it would look like to a relative observer in a typical branch}, and the argument was that it would look as if a random process had generated the records in accord with the quantum probabilities. In this sense, he was concerned with describing the properties of typical, not probable, branches. To get probabilities over branches, one would need to supplement Everett's account with something that allows one to move from what is true in a \emph{typical} branch to which branches are \emph{probable} or \emph{to be expected} together with an account of what such probabilities or expectations concern, and that would require auxiliary assumptions that go well beyond pure wave mechanics.

DeWitt and Graham seem to have wanted an explanation that addressed the question of what branch properties one should expect. But, if so, they clearly failed to deliver it. At the end of the paper where he criticizes Everett's choice of measure and introduces his new unbiased measure and partition of branches, Graham senses that he has not fully addressed the issue.
\begin{quote}
Thus we conclude that values of the relative frequency near [the standard predictions] will be found in the majority of Everett worlds of the apparatus and observer. If we assume our own world to be a ``typical'' one, then we may expect a human or mechanical observer to perceive relative frequencies in accord with the [standard predictions]. Why we should be able to assume our own world to be typical is, of course, itself an interesting question, but one that is beyond the scope of the present paper. (1973, 252)
\end{quote}
What Graham reports as being beyond the scope of the paper is why one should expect a principle of indifference to hold over worlds. But without that one cannot have anything like a deduction of standard quantum probabilities over branches, which begs the question as to the sense in which his partition is in any way preferable to Everett's in accounting for our experience.

The upshot is that whether one appeals to Everett's norm-squared coefficient measure~$m$ or to the unbiased indifference measure that Graham favors, one further needs to assume that the actual world is probably typical in order to get something like the standard quantum predictions from pure wave mechanics. The most direct way to accomplish this would be to add (1) a choice of partition, (2) a typicality measure over the partition, and (3) the assumption that the actual world is somehow randomly selected by probabilities determined by that measure to one's specification of the theory. But, again, there is good reason to suppose that Everett himself took it to be enough simply to establish that there is a sense in which the standard quantum statistics hold for a typical sequence of relative measurement records.\footnote{But, as discussed in Barrett (2015), one might require significantly more than Everett did from a satisfactory account of the standard quantum statistics.}

\section{The Everettian circle}

While Everett was free to use any typicality measure he wanted to make his argument, by this point one might suspect that there was nevertheless something ad hoc in his choice. The issue is perhaps clearest if one imagines that one's goal is somehow to deduce the probabilities that result from the standard collapse formulation of quantum mechanics as probabilities over branches, but the worry also holds if one considers Everett's more modest goal of just deducing the standard quantum statistics as typical.

In pure wave mechanics, a typical branch in the norm-squared coefficient measure~$m$ exhibits the standard quantum statistics. That is, if one were to randomly select a branch with probabilities determined by the norm-squared coefficient measure~$m$, one would expect to select a branch where the measurement records agree with the standard quantum statistics. But this fact is entirely unsurprising since a random selection of a branch with probabilities equal to this measure is precisely what the standard collapse dynamics accomplishes were one to perform a measurement on the final entangled state resulting from a series of correlating interactions. And the determinate records one expects to be produced by the collapse dynamics are what one means by the standard quantum statistics.

Put another way, since the measure Everett associated with each branch is just the standard quantum probability assigned to the sequence of measurement outcomes represented by that branch, if one were to randomly select a branch in that measure, one would be randomly selecting a branch with the standard quantum probabilities, which is precisely what the collapse dynamics does in the standard theory. Most simply, since Everett took the standard probability measure over branches to be his typicality measure, interpreting Everett's typicalities as probabilities, directly yields the standard probabilities over branches. The consequence is that if one modifies pure wave mechanics by stipulating that a particular branch is somehow randomly selected in measure~$m$ as actual, one's new theory simply recapitulates the explanation of quantum statistics given by the standard collapse theory. 

Further, since Everett's measure over branches is just the standard quantum probabilities over the sequences of measurement results represented by the branches, the fact that almost all branches in that measure exhibit the standard quantum statistics in the limit as one performs more measurements follows directly from the fact that the standard quantum statistics will almost certainly be exhibited by a branch selected from the state in the limit with the standard quantum probabilities. The only point of interest here is that one gets the same expected statistics applying the collapse dynamics \emph{once} after a series of linear measurement-like interactions as one gets applying it after \emph{each} of a series of linear measurement-like interactions.\footnote{A measurement-like interaction here is just one that correlates the pointer variable on one's measuring device with a property of the object system by means of the linear dynamics.}  But this is already an essential feature of the standard collapse theory. If this were not the case, then the standard theory would make entirely the wrong empirical predictions with the immediate consequence that it would be easy to determine which measurement-like interactions caused collapses and which did not.

The upshot is that if one understands Everett's typicality measure as giving the probability of each branch in fact being realized, then one gets the same predictions as the standard collapse theory and for precisely the same reason. One might try to avoid ending up where one started by taking Everett's measure to represent self-location probabilities rather than branch realization probabilities. On this view, while all of the branches are equally actual, one stipulates that the probability of \emph{finding oneself} in a particular branch is given by the Everett measure associated with that branch. Albert and Loewer's (1988) single- and many-minds theories provide a strategy for making sense of such self-location probabilities. On their account, which they present as a way of understanding Everett, an observer's physical state evolves on the standard unitary dynamics while Everett's measure~$m$ gives probabilities for the random evolution of the observer's mind(s). That is, instead of a collapse randomly selecting a branch to be realized, an observer's mind is randomly associated with the measurement records associated with a particular branch with probabilities given by~$m$.\footnote{While the Albert and Loewer single- and many-minds formulations of quantum mechanics are ontologically extravagant, they have the virtue of making it clear what self-location probabilities concern: they are probabilities that a particular mind will become associated with a particular branch.  If one wants a more metaphysically modest account of self-location probabilities, one has the challenge of being similarly clear regarding how they are to be understood. See Barrett (1999) for a discussion of the explanatory tradeoffs on accounts like Albert and Loewer's. In short, there is little to recommend their strategy beyond its clarity and that it makes the right probabilistic predictions and agrees with Everett's sense of what would happen if one could perform an interference measurement of one's friend and his measuring device and object system in a Wigner's Friend experiment.}

But shifting to self-location probabilities does not represent much of an advance. The standard collapse probabilities are probabilities for randomly realizing a branch corresponding to a particular set of determinate measurement records. Similarly, self-location probabilities are probabilities for randomly finding oneself associated with a branch that corresponds to a particular set of determinate measurement records. Hence, appropriating collapse probabilities for self-location yields the same quantum statistics as the standard collapse account and, again, for the same reason. In this sense, understanding Everett's measure~$m$ as providing self-location probabilities is arguably as ad hoc as understanding it as providing collapse probabilities.

That said, what matters more than the circularity of the choice of measure for the present argument is that making sense of self-location probabilities requires significant auxiliary assumptions concerning what self-location consists in and the relevance of quantum probabilities to where one finds oneself. The strength of the auxiliary assumptions required is concretely illustrated by what Albert and Loewer add to pure wave mechanics to get a clear account of self-location probabilities in single- and many-minds theories. More generally, in order to make sense of a random branch selection process one must make it clear precisely what the selection consists in and the relevance of the standard quantum probabilities to that selection.

One might concede the need for such auxiliary assumptions but insist that they are especially natural. After all, one might argue, Everett chose his typicality measure because it had special formal properties, not because it was the standard probability measure over measurement outcomes and hence was sure to yield the standard quantum statistics as typical. But, if one wants to deduce the standard quantum probabilities, it is difficult to see the special methodological virtue of (\emph{i}) pointing out that measure~$m$ has a number of suitable formal properties, then stipulating that~$m$ represents the standard quantum probabilities over (\emph{ii}) stipulating that~$m$ represents the standard quantum probabilities, then noting that it has a number of suitable formal properties. In each case, one is taking the measure that one already knows from the success of the standard collapse theory will yield the right empirical predictions and stipulating that it represents quantum probabilities. To insist that one's \emph{real reason} for the stipulation is purely formal is unconvincing when any other choice would entail the wrong predictions. This is not to say that one doesn't believe that one's aesthetic judgments concerning what is most natural given the formal structure of the theory are pure and consequently independent of pragmatic demands. Rather, empirical psychology provides good reason to suppose that we are particularly unreliable in judging how we form judgments in just such contexts.\footnote{While much has been done to study such phenomena since, the classic survey article is Nisbett and Wilson (1977).} Further, it unclear, at least to me, that judgments made here on purely aesthetic grounds are somehow more virtuous than those made from pragmatic necessity. Indeed, on reflection, it is unclear why one should expect aesthetic judgments to track truth at all here.

The point is not that one should use something other than~$m$ to assign quantum probabilities to possible measurement outcomes. If one insists on having the standard probabilistic predictions, then one might take the linear dynamics to describe the time-evolution of all physical systems and stipulate the standard collapse probabilities as something like self-location probabilities, filling in other significant background assumptions as one goes. But if so, one cannot seriously maintain that one is getting one probabilistic predictions from pure wave mechanics alone.

\section{Methodological Considerations}

It is unsurprising that reason alone cannot determine the standard quantum probabilities. In fact, it is entirely the other way around. Our experience of the behavior of quantum systems has proven to be deeply counterintuitive. One would never have guessed what the statistical properties of our quantum records would be before performing the appropriate experiments.

It is also unsurprising that one cannot derive the standard quantum probabilities from pure wave mechanics alone. The theory says nothing about probability. The unitary dynamics does not by itself require that one's measurement records exhibit any particular set of statistical properties. Indeed, pure wave mechanics does not even suggest a preferred typicality measure over branches or that any one result is realized at the expense of the others. While some measures may seem more natural than others given the empirical results we know we need to explain, that pure wave mechanics does not entail anything like a canonical measure can be seen in the disagreement between Everett and DeWitt and Graham concerning what measure is more natural. Everett knew that one might consider other measures. In the original version of his thesis he just wanted to argue that his choice of typicality measure was ``not as arbitrary as it appears'' (Barrett and Byrne 2012, 359).

If one wants anything like probabilities over branches from pure wave mechanics, one needs significant auxiliary assumptions. Further, if one insists on claiming that one can deduce quantum probabilities from one's physical theory alone, then honesty requires that one include these auxiliary assumptions in the specification of the theory.


The general methodological issue concerns what explanatory assumptions one has an obligation to include in one's specification of a theory. It would clearly be a mistake to include everything in one's theory that one might in fact use when giving theoretical explanations. One might implicitly or explicitly assume a good deal of mathematics, that one is not hallucinating, various boundary conditions, etc.\ without adding such assumptions to one's specification of the explanatory theory.

Closer to the present argument, while our physical theories typically say nothing explicit about \emph{experience}, we nevertheless often take them to make predictions about experience. By analogy one might argue that while pure wave mechanics says nothing about \emph{probability}, we might nevertheless take it to make probabilistic predictions.\footnote{David Wallace suggested this analogy in conversation.} Addressing this line of argument requires one to take a stand on when and how a theory explains.

We typically do not feel the need to add principles that stipulate the relationship between physical states and experience to our physical theories. Rather, we take a theory to make the right empirical predictions if it predicts determinate physical records on which one's experience might plausibly be taken to supervene. But, that said, when there is no straightforward supervenience relation at hand, we sometimes do explicitly stipulate a relationship between physical states and experience as a part of the specification of a physical theory. Indeed, that is just what Albert and Loewer's do in their single- and many-mind theories. The reason is that they want to account for the determinate experiences of observers when the physical state is the one predicted by the unitary dynamics and hence typically provides no clear candidate for something on which the experience of a determinate measurement outcome might supervene.\footnote{One might further argue that there being determinate outcomes is also a precondition for there being something one which probabilities over possible determinate outcomes might supervene.} By analogy, the suggestion is that one needs to add explicit assumptions regarding probabilities to pure wave mechanics precisely because the theory as it stands describes no chance events nor uncertainties on which quantum probabilities might supervene. Significantly, it is just that fact that gives the claim that one can derive quantum probabilities from pure wave mechanics alone its magical appeal.

We routinely appeal to auxiliary background assumptions that are not stated in the theory when we use a physical theory to explain. There is nothing wrong with that. But if one claims to be able to deduce the standard probabilities from a theory that says nothing about probability and describes no chance events or uncertainties on which quantum probabilities might supervene, then one needs to introduce probabilities and say what they concern and that will require auxiliary assumptions.  And if one insists that the standard quantum probabilities follow from one's theory alone, then one has a special responsibility to add the auxiliary assumptions that do the work to one's specification of the theory.

It is not always easy to say what principles need to be added to one's physical theory and what principles stand alone, but I can suggest a rule of thumb that seems to track at least some plausible explanatory demands in such situations. One may be comfortable not including a particular principle as part of the specification of one's physical theory if precisely that principle is both (1) compelling and genuinely useful in contexts independent of the physical theory and (2) clearly applicable to the physical theory given the usual justification for its use. Some principles, like Bayes' theorem, routinely pass the test. Other principles, like the principle of indifference, may never pass the test. To get the standard quantum probabilities from pure wave mechanics one needs a principle that connects branch amplitudes to probabilities. To satisfy condition (1) directly, one would need a principle that is both well motivated and useful independently of pure wave mechanics and ties quantum-mechanical amplitudes to probabilities. I can think of no such principle.

Introducing a typicality measure as a function of branch amplitudes might be seen as the first step in getting from amplitudes to probabilities. On this approach, getting the standard probabilistic predictions in the context of pure wave mechanics of the sort that DeWitt and Graham seem to have wanted would involve adding a notion of branch typicality \emph{then} adding a set of auxiliary assumptions that somehow tie the degree to which the value of a relative record is typical to the probability of that record being realized.

If one understands probability as a measure over possibilities where precisely one outcome is realized, one would need to explain what it might mean for a particular measurement record to be realized in pure wave mechanics. This is the strategy followed for most formulations of quantum mechanics. The standard formulation of quantum mechanics and GRW-type theories introduce a collapse dynamics that randomly realizes one branch at the expense of others, Albert and Loewer's (1988) single- and many-minds theories stipulate a dynamics that describes how an observer's experience is determined by a randomly selected branch, and hidden-variable theories like Bohmian mechanics add a parameter and an auxiliary dynamics that effectively selects a branch as having been realized with epistemic probabilities that explain the quantum statistics.\footnote{See Ghirardi, Rimini, and Weber (1986) for a description of the original GRW theory. See Bohm (1952) for an early description of Bohmian mechanics. Barrett (1999) provides further details regarding how probability works in the single- and many-minds theories and in Bohmian mechanics.} Of course, such additions produce theories that are less elegant and arguably more ad hoc than pure wave mechanics. But being worried that one's assumptions might be considered ad hoc does mean that one does not need them to get the standard quantum probabilities in each case. If one does not understand probability as a measure over possibilities where precisely one outcome is realized, then one needs to say how one understands probabilities in pure wave mechanics and the sense in which the theory can thereby be taken to account for our statistical experience.

Everett took collapse and hidden-variable theories to lack the simplicity and naturalness of pure wave mechanics because they added something to pure wave mechanics that it did not in fact need (1956, 152--7). But he purchases the elegance of pure wave mechanics with weak explanatory demands. Everett specified a notion of typicality such that the measurement records of a typical observer will exhibit the standard quantum statistics. And he took that to be enough to explain our statistical experience by rendering the theory empirically faithful.

Everett took his notion of typicality to put pure wave mechanics on a par with classical statistical mechanics in how it explains our experience as typical. Insofar as in pure wave mechanics all possible measurement records are explicitly taken to be actual, there is an important disanalogy between the two theories that he does not address. But the analogy is also important. While it does not allow us to infer probabilities or expectations, Everett's notion of typicality works with his notion of empirical faithfulness to explain our experience. In brief, pure wave mechanics is empirically faithful since one can find one's experience as typical in the model of the theory. The work done by Everett's notion of typicality, then, is in explaining experience. And the weakness of this explanation is indicated by the fact that it does not allow one to infer expectations without auxiliary assumptions that go well beyond pure wave mechanics as he characterized it.\footnote{See Barrett (2011a, 2015) for detailed discussions of Everett's notion of empirical faithfulness, the explanatory virtue that he took pure wave mechanics mechanics to exhibit, and a comparison of empirical faithfulness with stronger notions of empirical adequacy.}

\section{Conclusion}

Since pure wave mechanics says nothing whatsoever about probability, one must appeal to significant auxiliary assumptions to get quantum probabilities from the theory. Much of the work in deriving the standard quantum statistics from pure wave mechanics is done by one's choice of a measure and how to interpret it. The methodological point is that, given the special sort of explanatory role they play, such auxiliary assumptions should be included in one's specification of the theory one takes to explain quantum probabilities.

While there is no canonical way to distinguish between one's physical theory and one's beliefs and commitments more generally, there are some assumptions that one should include in the specification of one's theory. What these are depends on the explanatory context. The argument here is that one has a special obligation to add explicit assumptions regarding probabilities to pure wave mechanics because the theory does not mention of probabilities at all and describes no chance events nor uncertainties on which probabilities might supervene.

It is tempting to claim that pure wave mechanics alone explains the standard quantum probabilities precisely because it sounds like magic. And it would be. It is for this reason that one has a special explanatory burden of clearly specifying what one is relying on get probabilistic predictions if one wants them.

Everett took the standard quantum statistics to be satisfactorily explained by the fact that they are characteristic of a relative observer's relative records in a typical branch. His more modest explanatory goal still requires auxiliary assumptions, but they are weaker than what one would need to get quantum probabilities over branches. And weaker assumptions, unsurprisingly, yield a weaker sort of explanation.\footnote{I would like to thank David Wallace for discussions and Wayne Myrvold and three anonymous reviewers for helpful comments on an earlier version of this paper.}

\newpage

\begin{center}
\large{Bibliography}
\end{center}

\vspace{1cm}

\noindent

\vspace{.5cm}
\noindent
Albert, David, Z (2010) ``Probability in the Everett Picture,'' in Saunders, Barrett, Kent, and Wallace (eds) (2010), pp. 355--68.

\vspace{.5cm}
\noindent
Albert, David, Z (1986) ``How to Take a Photograph of Another Everett World'' \emph{Annals of the New York Academy of Sciences: New Techniques and Ideas in Quantum Measurement Theory} 480: 498--502.

\vspace{.5cm}
\noindent
Albert, David Z and Jeffrey A. Barrett (1995) ``On What It Takes to Be a World,'' \emph{Topoi}, 14(1): 35--37.

\vspace{.5cm}
\noindent
Albert, David and Loewer, Barry (1988) ``Interpreting the Many-Worlds Interpretation,'' Synthese. 77 (November): 195--213.

\vspace{.5cm}
\noindent
Barrett, Jeffrey A. (2016a) ``Quantum Worlds,'' \emph{Principia} 20(1): 45--60.

\vspace{.5cm}
\noindent
Barrett, Jeffrey A. (2016b) ``Typicality in Pure Wave Mechanics,'' \emph{Fluctuation and Noise Letters} 15(3).

\vspace{.5cm}
\noindent
Barrett, Jeffrey A. (2015) ``Pure Wave Mechanics and the Very Idea of Empirical Adequacy,'' \emph{Synthese} 192(10): 3071. 

\vspace{.5cm}
\noindent
Barrett, J. A. (2014) ``Description and the Problem of Priors,'' Erkenntnis 79(6): 1343--53.

\vspace{.5cm}
\noindent
Barrett, Jeffrey A. (2011a) ÒOn the Faithful Interpretation of Pure Wave MechanicsÓ \emph{British Journal for the Philosophy of Science} 62 (4): 693--709. 

\vspace{.5cm}
\noindent
Barrett, Jeffrey A. (2011b) ÒEverett's Pure Wave Mechanics and the Notion of WorldsÓ \emph{European Journal for Philosophy of Science} 1 (2):277--302.

\vspace{.5cm}
\noindent
Barrett, Jeffrey A. (1999) \emph{The Quantum Mechanics of Minds and Worlds}, Oxford: Oxford University Press.

\vspace{.5cm}
\noindent
Barrett, Jeffrey A. and Peter Byrne (eds) (2012) \emph{The Everett Interpretation of Quantum Mechanics: Collected Works 1955-1980 with Commentary}, Princeton: Princeton University Press.

\vspace{.5cm}
\noindent
Ben-Menahem, Y. and M.\ Hemmo (eds.) (2012) \emph{Probability in Physics}. Springer-Verlag: Berlin Heidelberg.

\vspace{.5cm}
\noindent
Bohm, David (1952) ``A Suggested Interpretation of Quantum Theory in Terms of `Hidden Variables','' Parts I and II, {\em Physical Review} 85: 166--179, 180--193.

\vspace{.5cm}
\noindent
DeWitt, Bryce S.\ (1971) ``The Many-Universes Interpretation of Quantum Mechanics'' in B.\ D.\ 'Espagnat (ed.) \emph{Foundations of Quantum Mechanics}. New York: Academic Press. Reprinted in DeWitt and Graham (1973) pp.\ 167--218.

\vspace{.5cm}
\noindent
DeWitt, Bryce S.\ and Neill Graham (eds.) (1973) \emph{The Many-Worlds Interpretation of Quantum Mechanics}. Princeton: Princeton University Press.

\vspace{.5cm}
\noindent
Deutsch, David (1999) ``Quantum Theory of Probability and Decisions,'' \emph{Proceedings of the Royal Society of London} A455: 3129--3137.

\vspace{.5cm}
\noindent
Dirac, P.\ A.\ M.\ (1930) \emph{The Principles of Quantum Mechanics}. Oxford: Oxford University Press.

\vspace{.5cm}
\noindent
Everett, Hugh III (1955a) ``Probability in Wave Mechanics'' Undated typescript addressed to John Wheeler with Wheeler's marginal notes. In Barrett and Byrne (eds) (2012, 64--70).

\vspace{.5cm}
\noindent
Everett, Hugh III (1955b) ``Objective vs Subjective Probability'' Undated typescript addressed to John Wheeler with Wheeler's marginal notes. In Barrett and Byrne (eds) (2012, 57--60).

\vspace{.5cm}
\noindent
Everett, Hugh III (1956) ``The Theory of the Universal Wave Function.'' In Barrett and Byrne (eds) (2012, 72--172).

\vspace{.5cm}
\noindent
Everett, Hugh III (1957) `` `Relative State' Formulation of Quantum Mechanics,'' Reviews of Modern Physics, 29: 454--462. In Barrett and Byrne (eds) (2012, 173--96).

\vspace{5mm}
\noindent
Farhi, E., J.\ Goldstone, S.\ Gutmann (1989) ``How Probability Arises in Quantum Mechanics,'' {\em Annals of Physics} 192(2) 368--382.

\vspace{5mm}
\noindent
Ghirardi, G.\ C., Rimini, A., and Weber, T.\ (1986) ``Unified dynamics for microscopic and macroscopic systems,'' \emph{Physical Review D} 34: 470--491. 

\vspace{5mm}
\noindent
Goldstein, Sheldon (2012) ``Typicality and Notions of Probability in Physics,'' in \emph{Probability in Physics}, edited by Y. Ben-Menahem and M. Hemmo, 59--71.

\vspace{.5cm}
\noindent
Graham, N.\ (1973) ``The Measurement of Relative Frequency,'' in DeWitt and Graham (eds) (1973).

\vspace{.5cm}
\noindent
Greaves, Hillary (2007) ``Probability in the Everett Interpretation'' \emph{Philosophy Compass} 2/1:109--128. 10.1111/j.1747-9991.2006.00054.x

\vspace{5mm}
\noindent
Hartle, J.\ B.: (1968) ``Quantum Mechanics of Individual Systems,'' {\em American Journal of Physics} 36(8) 704--12.

\vspace{5mm}
\noindent
Kent, Adrian (2010) ``One World Versus Many: The Inadequacy of Everettian Accounts of Evolution, Probability, and Scientific Confirmation,'' in Saunders et al. (eds) (2010).

\vspace{5mm}
\noindent
Nisbett, Richard E., Wilson, Timothy D. (1977). ``Telling More than We Can Know: Verbal Reports on Mental Processes" \emph{Psychological Review} 84(3): 231--59.

\vspace{.5cm}
\noindent
Saunders, Simon; Jonathan Barrett; Adrian Kent; David Wallace (eds) (2010) \emph{Many Worlds?: Everett, Quantum Theory, and Reality}, Oxford: Oxford University Press.

\vspace{.5cm}
\noindent
Saunders, Simon (2010) ``Chance in the Everett Interpretation'' in Saunders et al. (eds) (2010) pp. 181--205.

\vspace{.5cm}
\noindent
Sebens, Charles T.\ and Sean M. Carroll (2016) ``Self-Locating Uncertainty and the Origin of Probability in Everettian Quantum Mechanics,'' \emph{British Journal for the Philosophy of Science} doi: 10.1093/bjps/axw004. First published online: July 5, 2016

\vspace{.5cm}
\noindent
Vaidman, Lev (2016) ``Many-Worlds Interpretation of Quantum Mechanics,'' \emph{The Stanford Encyclopedia of Philosophy} Edward N. Zalta (ed).
Accessed 4 October 2016.

\vspace{.5cm}
\noindent
Vaidman, Lev (2012) ``Probability in the Many-Worlds Interpretation of Quantum Mechanics,'' in Yemima Ben-Menahem and Meir Hemmo (eds.), \emph{Probability in Physics (The Frontiers Collection XII)} Berlin Heidelberg: Springer, pp. 299--311.

\vspace{.5cm}
\noindent
van Fraassen, Bas (1989) \emph{Laws and Symmetry} Oxford: Clarendon Press.

\vspace{.5cm}
\noindent
von Neumann, J.\ (1955) \emph{Mathematical Foundations of Quantum Mechanics}. Princeton: Princeton University Press. Translated by R. Beyer from \emph{Mathematische
Grundlagen der Quantenmechanik}. Berlin: Springer (1932).

\vspace{.5cm}
\noindent
Wallace, David (2012) \emph{The Emergent Multiverse: Quantum Theory according to the Everett Interpretation}, Oxford: Oxford University Press.

\vspace{.5cm}
\noindent
Wallace, David (2010a) ``Decoherence and Ontology,'' in Saunders et al (2010) pp. 53--72.

\vspace{.5cm}
\noindent
Wallace, David (2010b) ``How to Prove the Born Rule,'' in Saunders et al (2010) pp. 227--63.

\vspace{.5cm}
\noindent
Wigner, Eugene (1961) ``Remarks on the Mind-Body Problem'', in I. J. Good (ed.), \emph{The Scientist Speculates}, New York: Basic Books, pp.\ 284--302.

\vspace{.5cm}
\noindent
Zurek, Wojciech H. (2005) ``Probabilities from Entanglement, Born's Rule $p_k = |\psi_k|^2$ from Envariance,'' \emph{Physical Review A} 71(5): 052105.

\end{document}